\documentclass{mn2e}

\usepackage{graphicx}                    
\usepackage{colorrgb}
\usepackage{times}
\usepackage{bbold}

\def\rot{\mathop{\rm rot}\nolimits}
\def\div{\mathop{\rm div}\nolimits} 
\def\gsim{\lower.4ex\hbox{$\;\buildrel >\over{\scriptstyle\sim}\;$}} 
\def\lsim{\lower.4ex\hbox{$\;\buildrel <\over{\scriptstyle\sim}\;$}} 
\def\q{\qquad}
\def\beg{\begin{eqnarray}}
\def\ende{\end{eqnarray}}
\renewcommand{\vec}[1]{\mbox{\boldmath $#1$}}

\title[Toroidal magnetic fields and differential rotation]
{Destabilisation of hydrodynamically stable rotation laws
by azimuthal magnetic fields}

\author[G. R\"udiger et al.]
{G\"unther R\"udiger$^1$\thanks{E-mail: gruediger@aip.de},  Rainer Hollerbach$^2$, Manfred Schultz$^1$ and Detlef Elstner$^1$\\
$^1$Astrophysikalisches Institut Potsdam, An der Sternwarte 16, 
D-14482 Potsdam, Germany
\\   
$^2$Department of Applied Mathematics, University of 
Leeds, Leeds, LS2 9JT, UK}
\begin{document}
\maketitle
\label{firstpage}
\begin{abstract}
We consider the effect of toroidal magnetic fields on hydrodynamically
stable Taylor-Couette differential rotation flows.  For current-free
magnetic fields a nonaxisymmetric $m=1$ magnetorotational instability
arises when the magnetic Reynolds number exceeds $O(100)$.  We then consider
how this `azimuthal magnetorotational instability' (AMRI) is modified
if the magnetic field is not current-free, but also has an associated electric
current throughout the fluid.  This gives rise to current-driven Tayler
instabilities (TI) that exist even without any differential rotation at
all.  The interaction of the AMRI and the TI is then considered when
both electric currents and differential rotation are present simultaneously.
The magnetic Prandtl number Pm turns out to be crucial in this case.
Large Pm have a destabilizing influence, and lead to a smooth transition
between the AMRI and the TI.  In contrast, small Pm have a stabilizing
influence, with a broad stable zone separating the AMRI and the TI.
In this region the differential rotation is acting to stabilize
the Tayler instabilities, with possible astrophysical applications (Ap
stars).  The growth rates of both the AMRI and the TI are largely
independent of Pm, with the TI acting on the timescale of a
single rotation period, and the AMRI slightly slower, but still on the
basic rotational timescale.  The azimuthal drift timescale is $\sim20$
rotations, and may thus be a (flip-flop) timescale of stellar activity
between the rotation period and the diffusion time. 
\end{abstract}


\section{Introduction}
We consider the problem of how linear hydrodynamic and
magnetohydrodynamic instabilities interact in a simple rotating shear
flow, the familiar Taylor-Couette flow between concentric cylinders.
In the purely hydrodynamic problem, the Rayleigh criterion states that
an ideal flow is stable against axisymmetric perturbations when the
specific angular momentum increases outwards
\beg
\frac{\rm d}{{\rm d}R}(R^2\Omega)^2 > 0,
\label{ray}
\ende
where $\Omega$ is the angular velocity, and cylindrical coordinates
($R$, $\phi$, $z$) are used.  Viscosity has a stabilizing effect,
so that a Taylor-Couette flow that violates (\ref{ray}) becomes unstable
only if the angular velocity of the inner cylinder  (that is, its
Reynolds number) exceeds some critical value. 

If a uniform axial magnetic field is included, a new type of instability
arises, the  magnetorotational instability (MRI).  The Rayleigh
criterion is then replaced by
\beg
\frac{\rm d}{{\rm d}R}\Omega^2 > 0.
\label{mri}
\ende
That is, the requirement for stability is that the angular velocity
itself increases outward, rather than the angular momentum.  This is
more stringent than (\ref{ray}), so a flow may be hydrodynamically
stable, but magnetohydrodynamically unstable.  And again, viscosity
and/or magnetic diffusivity have a stabilizing effect; for small
magnetic Prandtl numbers it is the magnetic Reynolds number that must
exceed a critical value for instability to occur (R\"udiger, Schultz
\& Shalybkov 2003).  Hollerbach \& R\"udiger (2005) considered how the
magnetorotational instability is modified if an azimuthal field $B_\phi
\propto R^{-1}$ is added, and found that the relevant parameter is then
the ordinary Reynolds number $\rm Re$, rather than the magnetic Reynolds
number $\rm Rm$.

In this work we will consider instabilities of purely azimuthal magnetic
fields, without any axial field being present.  We begin by showing that
even $B_\phi\propto R^{-1}$ by itself supports a magnetorotational instability.
This new type of MRI is nonaxisymmetric, having azimuthal wavenumber $m=1$,
but otherwise shares all the characteristics of the classical, axisymmetric
MRI in an axial field. These results are presented in section 3.

We then extend the choice of imposed field to be of the form $B_\phi=a_B R
+ b_B R^{-1}$.  The part $b_B R^{-1}$ corresponds to an axial current only
within the inner cylinder $R<R_{\rm{in}}$, so current-free within the fluid,
but the part $a_B R$ corresponds to a uniform axial current density $J_z=2a_B$
everywhere within $R<R_{\rm{out}}$. Taking $B_\phi=a_B R + b_B R^{-1}$ instead
of just $B_\phi=b_B R^{-1}$ has profound implications, well beyond simply
having a somewhat different radial profile.  In particular, if there are no
electric currents flowing within the fluid ($a_B=0$), the only source of
energy to drive instabilities is the differential rotation; the magnetic
field merely acts as a catalyst, not as a source of energy.  For $\rm Re=0$
no instabilities are thus possible, regardless of how strong the magnetic
field is.

In contrast, if there are electric currents flowing within the fluid ($a_B
\neq0$), this yields a new source of energy to drive purely magnetic instabilities, that may exist even at $\rm Re=0$.  Michael (1954) and
Velikhov (1959) considered axisymmetric magnetic instabilities and derived
the stability criterion
\beg
\frac{\rm d}{{\rm d}R}\left( \frac{B_\phi}{R} \right)^2 < 0.
\label{diss}
\ende
Tayler (1973) included nonaxisymmetric disturbances and showed that for an
ideal fluid the necessary and sufficient condition for stability is
\beg
 \frac{\rm d}{{\rm d}R}( R B_\phi^2) < 0.
\label{tay}
\ende
A {\em uniform} field would therefore be axisymmetrically stable, but
nonaxisymmetrically unstable, with $m=1$ being the most unstable mode.
(The profile $B_\phi\propto R^{-1}$ is stable according to both (\ref{diss})
and (\ref{tay}), in agreement with the discussion above, that for such a
current-free profile there is simply no source of energy to drive purely
magnetic instabilities.)

In section 4 we then choose $a_B$ and $b_B$ such that $B_\phi$ is as uniform
as possible, with the same values at $R_{\rm in}$ and $R_{\rm out}$.  We
consider how the resulting Tayler instabilities (TI) interact with the
azimuthal magnetorotational instabilities (AMRI) presented in section 3.
We find that the magnetic Prandtl number plays a key role: if $\rm Pm\gsim10$
the AMRI and the TI are smoothly connected to one another, but if $\rm Pm
\lsim1$ they are disconnected, with a region of stability separating them.

The results presented here have important astrophysical implications,
since the simultaneous existence of differential rotation and toroidal
magnetic fields is characteristic of almost all celestial bodies.
Prominent examples of differentially rotating objects with large magnetic
Prandtl number are protogalaxies (without supernova explosions) and
protoneutron stars (PNS). Even rather weak toroidal fields should be
unstable in these objects.  In contrast, the radiative zones of stars have
$\rm Pm<10^{-2}$. In this case we will see that the toroidal field is
{\em stabilized}, as long as it is not too strong.  Finally, in the limit of
no differential rotation, the pure Tayler instability is independent of
$\rm Pm$.

\section{Basic equations}

We consider a viscous, electrically conducting, incompressible fluid between
two rotating infinite cylinders, in the presence of an azimuthal magnetic
field. The equations of the problem are
\begin{eqnarray}
\lefteqn{\frac{\partial \vec{U}}{\partial t} + (\vec{U}\cdot\nabla)\vec{U} =
-\frac{1}{\rho} \nabla P + \nu \Delta \vec{U} + 
\frac{1}{\mu_0\rho} {\rot \vec{B}} \times \vec{B},}
\nonumber \\
\lefteqn{\frac{\partial \vec{B}}{\partial t}= {\rot}\ (\vec{U} \times \vec{B})
+ \eta \Delta\vec{B},}
\label{mhd}
\ende
and
\beg
{\div}\ \vec{U} = {\div}\ \vec{B} = 0,
\label{mhd1}
\ende
where $\vec{U}$ is the velocity, $\vec{B}$ the magnetic field, $P$ the 
pressure, $\nu$ the kinematic viscosity, and $\eta$ the magnetic
diffusivity.  Equations (\ref{mhd}) yield the basic state solution 
\beg
U_R=U_z=B_R=B_z=0, 
\ende
and
\beg
U_\phi=R\Omega=a_\Omega R+\frac{b_\Omega}{R}, \q
B_\phi=a_B R+\frac{b_B}{R}.
\label{basic}
\ende
$a_\Omega$ and $b_\Omega$ are given by
\beg
a_\Omega=\Omega_{\rm{in}}\frac{\hat \mu_\Omega-{\hat\eta}^2}{1-{\hat\eta}^2}, \q
b_\Omega=\Omega_{\rm{in}} R_{\rm{in}}^2 \frac{1-\hat\mu_\Omega}{1-{\hat\eta}^2},
\ende
where
\begin{equation}
\hat\eta=\frac{R_{\rm{in}}}{R_{\rm{out}}}, \q
\hat\mu_\Omega=\frac{\Omega_{\rm{out}}}{\Omega_{\rm{in}}}.
\label{mu}
\end{equation}
$\Omega_{\rm in}$ and $\Omega_{\rm out}$ are the imposed rotation rates of
the inner and outer cylinders, with radii $R_{\rm in}$ and $R_{\rm out}$.

In contrast to $U_\phi$, where $a_\Omega$ and $b_\Omega$ are merely derived
quantities ($\Omega_{\rm in}$ and $\Omega_{\rm out}$ being the fundamental
quantities), for $B_\phi$, $a_B$ and $b_B$ themselves are the fundamental
quantities; as previously noted, $a_B R$ corresponds to a uniform axial
current everywhere within $R<R_{\rm out}$, and $b_B R^{-1}$ corresponds to
an additional current only within $R<R_{\rm in}$.  In analogy with
$\hat\mu_\Omega$, it is useful though to define the quantity
\begin{equation}
\hat\mu_B=\frac{B_{\rm{out}}}{B_{\rm{in}}}
  =\frac{a_B R_{\rm out}+b_B R_{\rm out}^{-1}}
        {a_B R_{\rm in }+b_B R_{\rm in }^{-1}},
\end{equation}
measuring the variation in $B_\phi$ across the gap.

In section 3 we consider the field $B_\phi\propto R^{-1}$, so $a_B=0$ and
$b_B=1$.  In contrast, in section 4 we include nonzero $a_B$, and adjust it
so that $\hat\mu_B=1$, so the toroidal field profile is as uniform as
possible (see also Cally 2003).  Taking $R_{\rm in}=1$ and $R_{\rm out}=2$,
this corresponds to setting $a_B=1/3$ and $b_B=2/3$; $B_\phi$ then varies
by less than 6\% across the gap.

We are interested in the linear stability of the solution (\ref{basic}).
The perturbed state of the flow is described by
\beg
u_R, \; R\Omega+u_\phi , \; u_z, \; b_R,  \; B_\phi+b_\phi, \; b_z.
\ende
Developing the disturbances into normal modes, solutions of the linearized
equations are considered in the form
\beg
F=F(R){\rm{exp}}\bigl[{\rm{i}}(kz+m\phi+\omega t)\bigr],
\label{nmode}
\ende
where $F$ is any of the perturbation quantities. 

The magnetic Prandtl number (Pm), the Hartmann number (Ha) and the
Reynolds number (Re) are the dimensionless numbers of the problem,
\beg
{\rm{Pm}} = \frac{\nu}{\eta}, \q\ 
{\rm{Ha}}=\frac{B_{\rm{in}} R_0}{\sqrt{\mu_0 \rho \nu \eta}},  \q\ 
{\rm{Re}}=\frac{\Omega_{\rm{in}} R_0^2}{\nu},
\label{pm}
\ende
where $R_0=(R_{\rm{in}}(R_{\rm{out}}-R_{\rm{in}}))^{1/2}$
is taken as the unit of length.  We have used $R_0^{-1}$ as the unit of
the wave number, $\eta/R_0$ as the unit of the velocity fluctuations,
$\Omega_{\rm{in}}$ as the unit of frequencies, and $B_{\rm{in}}$ as the
unit of the magnetic field fluctuations.  Where appropriate, we will also
use the magnetic Reynolds number
\beg\rm Rm= Pm \cdot Re\ende
and the Lundquist number
\beg\rm S= \sqrt{Pm} \cdot Ha.\ende

A set of ten boundary conditions is needed to solve the equations, namely
no-slip
\beg
u_R=u_\phi=u_z=0
\label{ubnd}
\ende
for the flow, and perfectly conducting
\beg
{\rm d}b_\phi/{\rm d}R + b_\phi/R = b_R = 0
\label{bcond}
\ende
for the field, both sets being applied at both $R_{\rm{in}}$ and
$R_{\rm{out}}$.  If the exterior regions were taken to be insulators,
the boundary conditions on the field would be different (e.g. R\"udiger,
Schultz \& Shalybkov 2003; Shalybkov 2006), but here we will consider only conducting
boundary conditions.

The system of linearized equations and associated boundary conditions
then constitutes a one-dimensional linear eigenvalue problem, solved by
finite-differencing in radius  as in R\"udiger et al. (2005).  Typically
around 100 grid points were used, and all results were checked to ensure
that they were fully resolved.

\section{The azimuthal magnetorotational instability (AMRI)}

\begin{figure}
\includegraphics[width=8.5cm,height=7.0cm]{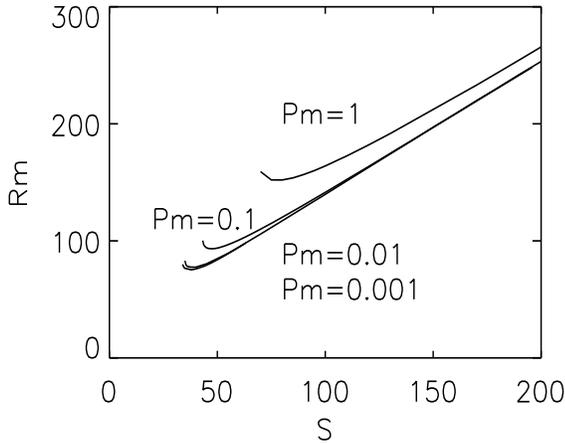}
\caption{\label{fig1} The marginal stability domains for the $m=1$ azimuthal 
magnetorotational instability (AMRI)  with $\hat\eta=0.5$ conducting cylinders.
The outer cylinder rotates with 50\% of the rotation rate of the inner one,
$\hat\mu_\Omega=0.5$; the magnetic field between the cylinders is current-free, 
$\hat\mu_B=0.5$.  The curves are marked with their magnetic Prandtl numbers.}
\end{figure}

The  rotation laws in stars and galaxies are hydrodynamically stable. Such 
rotation laws can be modeled by Taylor-Couette containers with rotating outer 
cylinder. More precisely, the rotation of the outer cylinder must fulfill the 
Rayleigh condition, $\hat\mu>\hat\eta^2$. The containers considered here have 
$\hat\eta=0.5$, so that a flow with $\hat\mu=0.5$ is clearly beyond the 
Rayleigh limit where the hydrodynamic instability disappears. Such a rotation 
law, however, can become unstable against nonaxisymmetric disturbances
under the presence of a current-free toroidal field. Figure \ref{fig1}
shows stability curves for $m=1$, the only mode that appears to become
unstable.  We see how a magnetorotational instability exists that is
remarkably similar to the classical MRI with a purely axial field.
In particular, as ${\rm Pm}\to0$, the relevant parameters are also $\rm Rm$
and $\rm S$, with the MRI arising if ${\rm Rm}\gsim80$, and ${\rm S}\approx40$
yielding the lowest value of $\rm Rm_c$.  The specific numbers are roughly an
order of magnitude greater than for the axisymmetric MRI in the axial field
(cf. R\"udiger, Schultz \& Shalybkov 2003), but the basic scalings, and even
the detailed shape of the instability curves, are identical.

Figure \ref{fig11} shows the real and imaginary parts of ${\rm i}\omega$ in the
unstable regime.  Remembering that time has been scaled by
$\Omega_{\rm in}^{-1}$, we see that we obtain growth rates as large as
$0.05\Omega_{\rm in}$.  So again, while the particular number 0.05 is about
an order of magnitude smaller than for the axisymmetric MRI in the axial
field, this nonaxisymmetric MRI is clearly also growing on the basic
rotational timescale.
\begin{figure}
\vbox{
\includegraphics[scale=0.45]{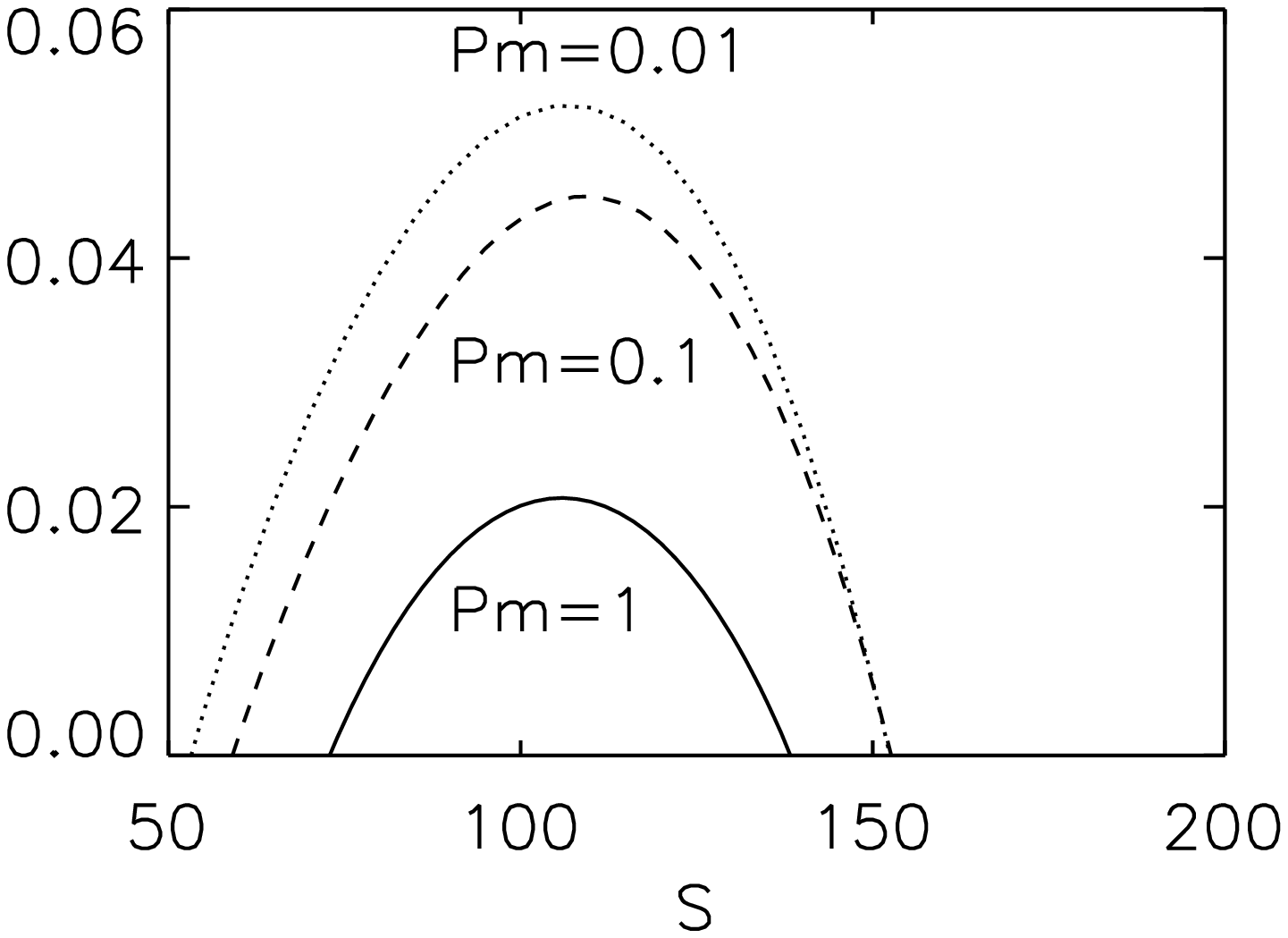}
\includegraphics[scale=0.45]{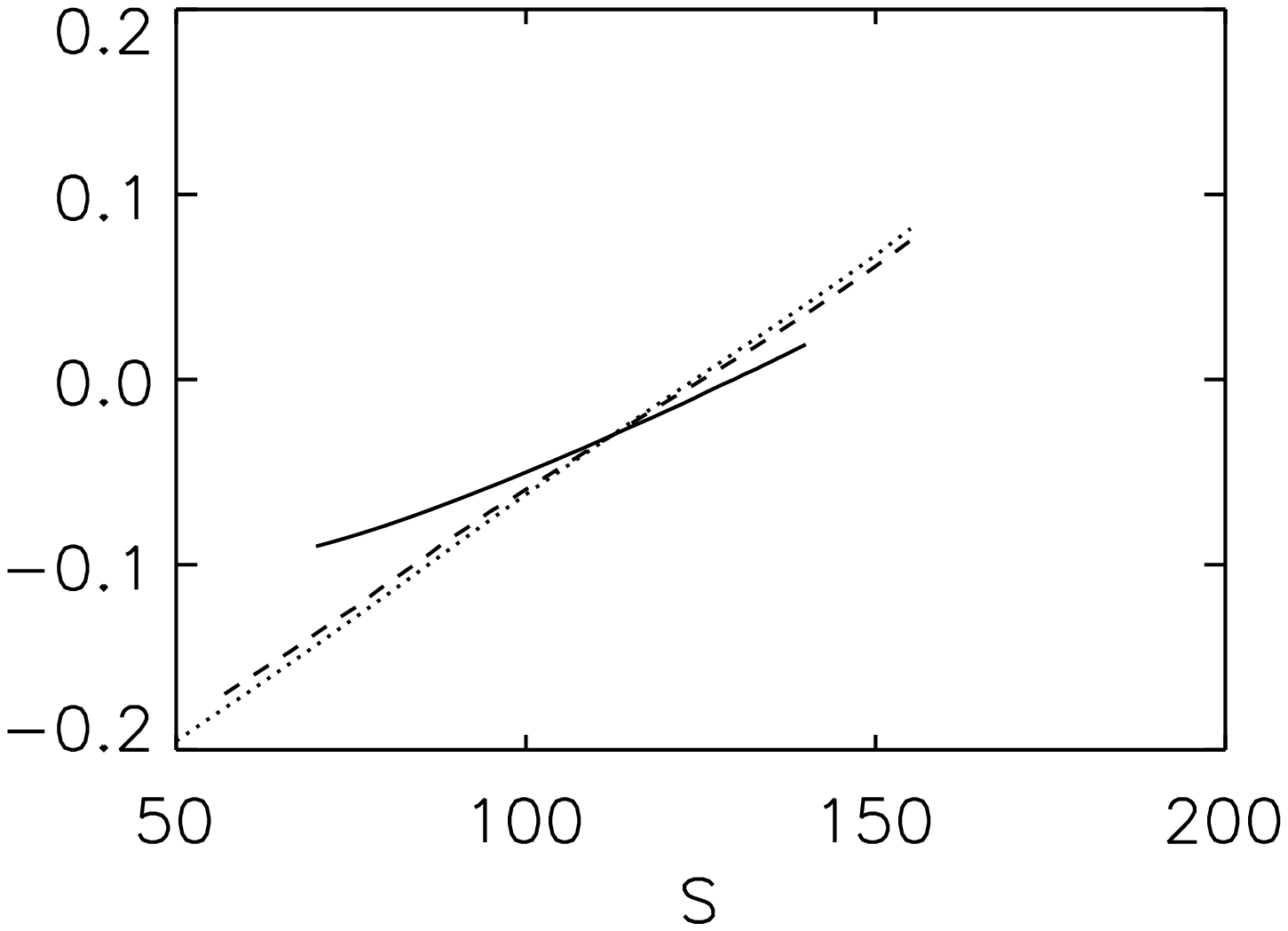}}
\caption{The growth rates (top) and azimuthal drift rates (bottom), as
functions of $\rm S$, with $\rm Rm$ fixed at 200.}
\label{fig11}
\end{figure}

To understand why this nonaxisymmetric MRI exists even for a purely
toroidal field ${\vec B}_0$, for which it is known that the axisymmetric MRI
fails, we need to consider the $r$ and $\phi$ components of the induction
equation,
\begin{eqnarray}
\lefteqn{{\rm Rm}\,{\rm i}\omega b_R=\Delta b_R - R^{-2}b_R - 2{\rm i}mR^{-2}b_\phi+}\nonumber\\
&& \qquad\quad+{\rm i}mR^{-2}u_R-{\rm Rm}\,{\rm i}m\Omega\,b_R,
\end{eqnarray}
\begin{eqnarray}
\lefteqn{{\rm Rm}\,{\rm i}\omega b_\phi=\Delta b_\phi - R^{-2}b_\phi + 2{\rm i}mR^{-2}b_R+}\nonumber\\
\lefteqn{ \ \ \  +{\rm i}mR^{-2}u_\phi+2R^{-2}u_R-{\rm Rm}\,{\rm i}m\Omega\,b_\phi
 + {\rm Rm}\,R\frac{{\rm d}\Omega}{{\rm d}R}\, b_R.}
\end{eqnarray}
In particular, note that for $m=0$, $b_R$
completely decouples from everything else, and inevitably decays away.
Without $b_R$ though, the MRI cannot proceed, as it relies on the term
$R({{\rm d}\Omega}/{{\rm d}R})b_R$.  In contrast, for $m=1$, $b_R$ is
coupled both to $b_\phi$, coming from $\Delta{\vec b}$, and to $u_R$, from
$\rot(\vec{ u}\times \vec{B}_0)$.  And once $b_R$ is coupled to the rest of the
problem, the term $R({{\rm d}\Omega}/{{\rm d}R})b_R$ then allows the MRI
to develop.  A detailed examination of the structure of the solutions shows
that all three components of both $\vec u$ and $\vec b$ are indeed present.

We emphasize also that this indeed is a magnetorotational instability, and
not a pinch, Tayler, or other current-driven instability.  Once again, for
a current-free field all current-driven instabilities are excluded {\it a
priori}.  What we have here is that a magnetic field, which by itself would
be stable at any amplitude, acts as a catalyst and destabilizes a
hydrodynamically stable differential rotation, just as in the classical MRI.
We note though that there is no hope of realizing this AMRI in the laboratory;
not only is the critical magnetic Reynolds number rather large, but beyond
that, achieving a sufficiently strong toroidal field would require enormously
large electric currents, beyond what could reasonably be imposed.

\begin{figure}
\vbox{
\includegraphics[width=7cm,height=4.3cm]{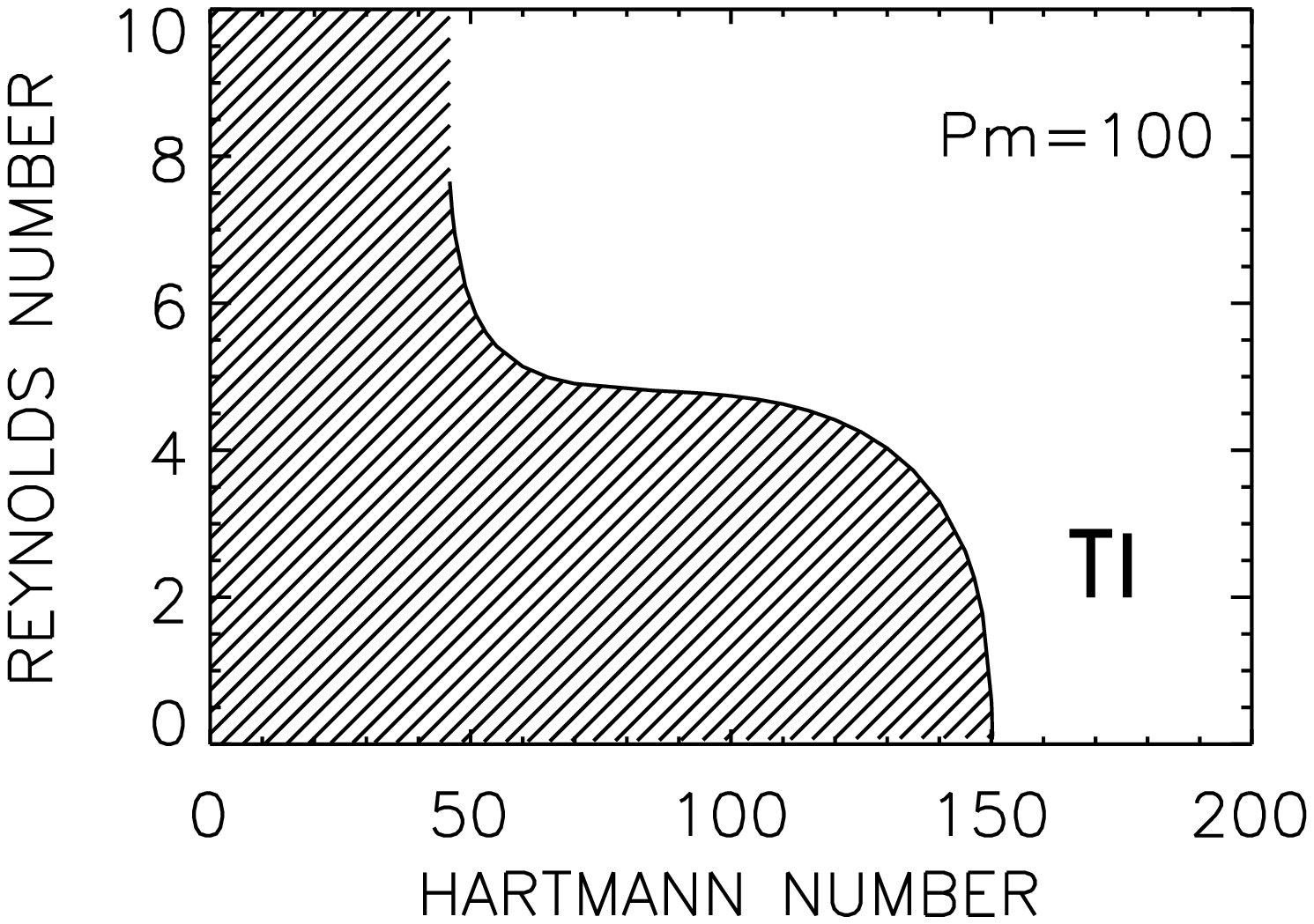}
\includegraphics[width=7cm,height=4.3cm]{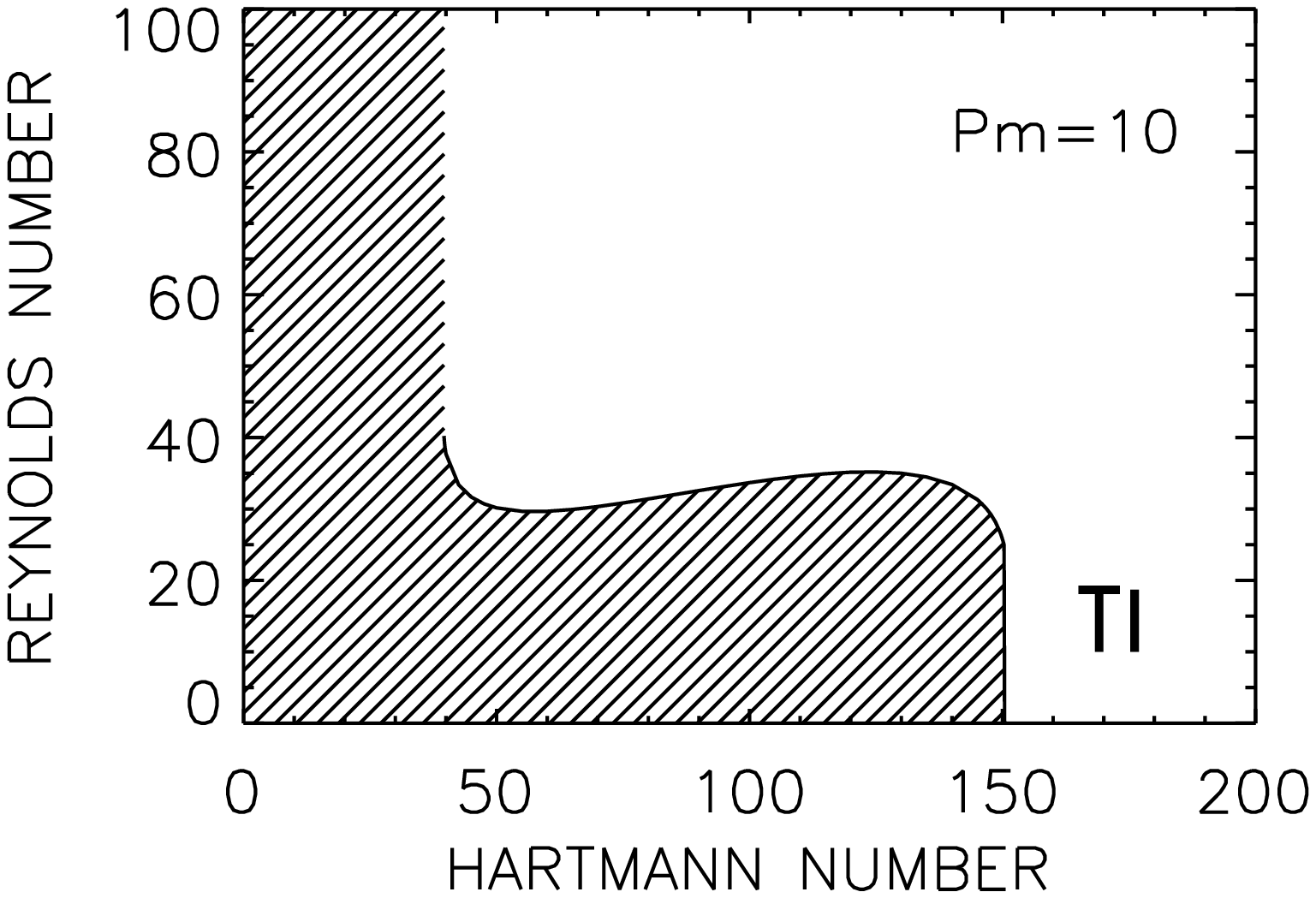}
\includegraphics[width=7cm,height=4.3cm]{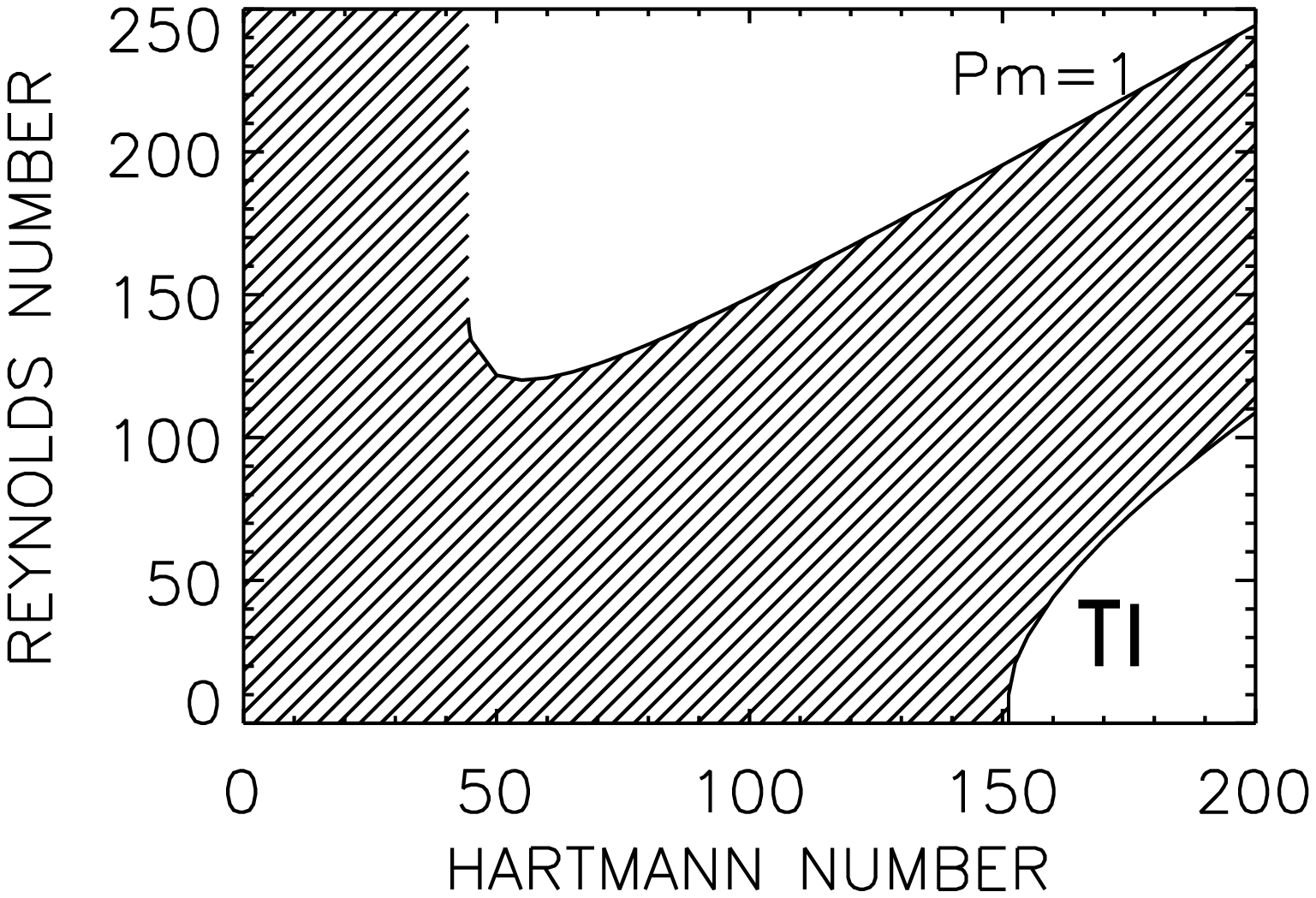}
\includegraphics[width=7cm,height=4.3cm]{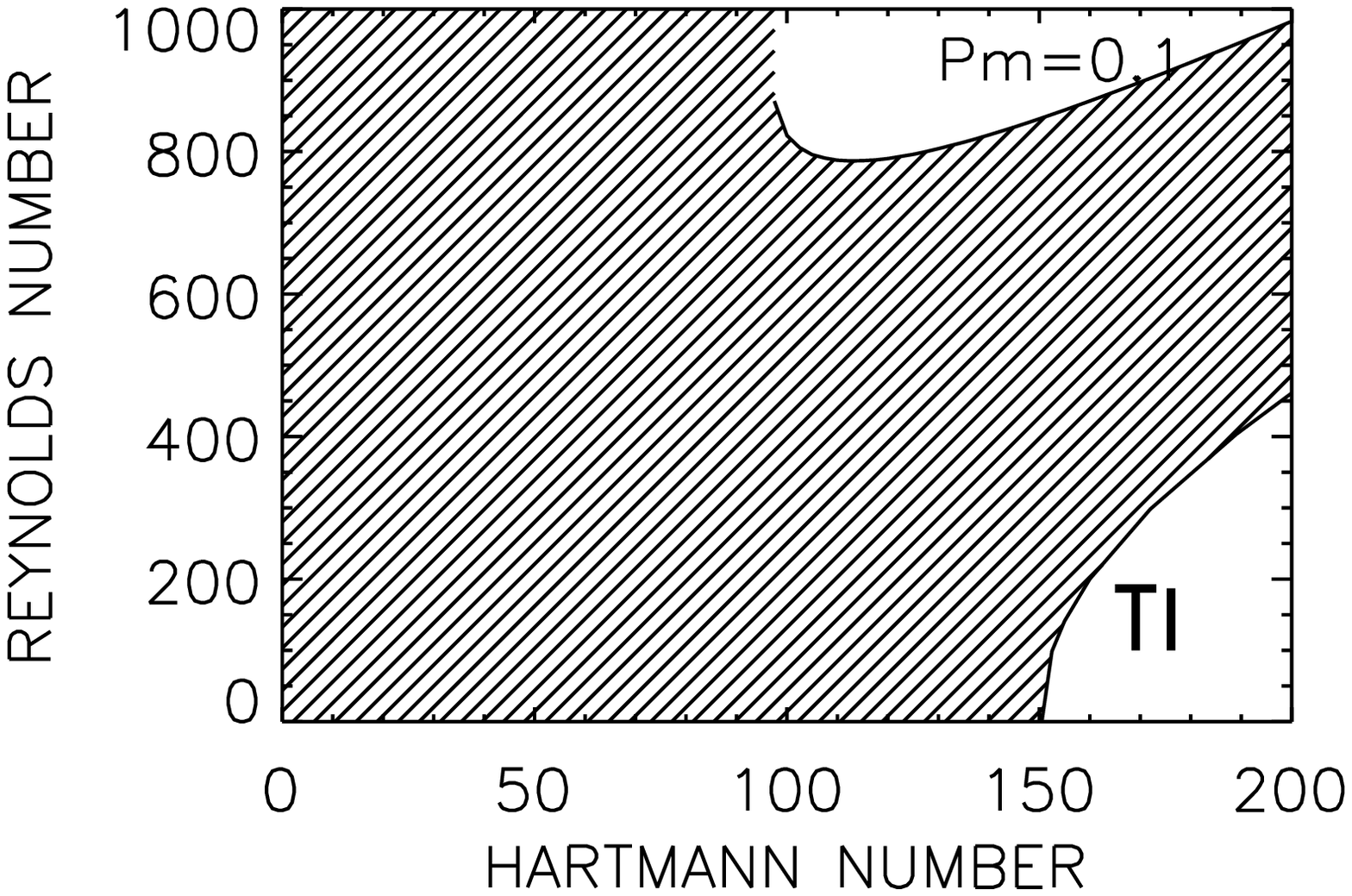}}
\caption{\label{fig2}
The marginal stability domains (hatched) for the $m=1$ TI and AMRI together,
for $\hat\eta=0.5$ conducting cylinders.  The outer cylinder rotates with 50\%
of the rotation rate of the inner one, $\hat\mu_\Omega=0.5$; the magnetic
field between the cylinders is as uniform as possible, $\hat\mu_B=1$.  The
panels are marked with their magnetic Prandtl numbers.}
\end{figure}

\section{The Tayler instability (TI)}

Having demonstrated that the current-free field $B_\phi=R^{-1}$ yields this
nonaxisymmetric azimuthal magnetorotational instability, we next wish to
consider what effect including currents within the fluid ($a_B\neq0$) has.
There are two questions one might wish to address here.  First, how robust
is the AMRI in this case; does it continue to exist at all, and if so, are
the growth and drift rates much the same?  Second, as discussed in the
introduction, if there are currents flowing within the fluid, then for
sufficiently great Hartmann numbers Tayler instabilities exist (here
$\rm Ha_{\rm Tayler}\simeq 150$), which also turn out to be $m=1$, and
could thus be expected to interact in some significant way with the AMRI.
In this section we therefore consider the nature of this interaction between
the AMRI and the TI, and in particular how it depends on the magnetic Prandtl
number Pm.

Generally, as one can read from Fig. \ref{fig2} fast rotation is stabilizing and strong toroidal fields are stabilizing (Pitts \& Tayler 1985). However, large magnetic Prandtl numbers also prove to be  strongly destabilizing (Fig. \ref{fig2}, top).  The right half of the AMRI branch, the one for strong magnetic fields,
disappears completely.  Instead, the critical Rm decreases monotonically
with Ha until eventually $\rm Ha_{\rm Tayler}\simeq 150$ is reached. The
transition from the AMRI to the TI is smooth, without any striking features.

However, for $\rm Pm\lsim 1$ the stable right branch of the AMRI in
Fig. \ref{fig1} reappears. There is always a domain between the AMRI and the
TI where the flow is {\em stable}, despite the large magnetic fields,
$\rm Ha>Ha_{\rm Tayler}$), that is, in a regime where the field without a
differential rotation would be unstable.  Roughly speaking, according to
the results  in Fig. \ref{fig2} (bottom), the TI exists for sufficiently
strong magnetic field, i.e.
\begin{equation}
{\rm S}\gg {\rm Rm}
\label{gg}\end{equation}
and the AMRI exists for sufficiently weak magnetic field, i.e.
\begin{equation}
{\rm S} \ll {\rm Rm}
\label{ll}\end{equation}
(for $\rm Pm \lsim 1$).
With  the magnetic velocity $v_{\rm A} =B_\phi/\sqrt{\mu_0\rho}$ and the linear rotation speed $U_\phi=R_{\rm in} \Omega_{\rm in}$ then
\begin{equation}
\frac{v_{\rm A}}{U_\phi} \gg \frac{\Delta R}{R}
\label{ggg}\end{equation}
for TI and 
\begin{equation}
\frac{v_{\rm A}}{U_\phi} \ll \frac{\Delta R}{R}
\label{lll}\end{equation}
 for AMRI.  Here $\Delta R/R$ is the fractional layer thickness which is of order unity in our  model   while for the solar tachocline its value is about 0.05.  The AMRI and TI are separated by an extended stable domain where
$v_{\rm A} / U_\phi \approx \Delta R/R$.

For small Pm and for sufficiently high Reynolds number  a second  (smaller) critical Hartmann number exists so that  depending on the initial conditions two different stable solutions are possible. A solution with growing amplitude (e.g. dynamo-induced) becomes unstable already at $\rm Ha\simeq 50$ while a decaying magnetic field (e.g. after collapse) already becomes stable  at $\rm Ha > 150$. 

For our special magnetic-field profile ($\hat \mu_B=1$) and for sufficiently
large Reynolds numbers, the lower critical Hartmann number moves from 150 for 
small Pm to about 50 for large Pm (see Fig. \ref{fig2}).  Large Pm are
destabilizing, and small Pm are stabilizing (Kurzweg 1963).  However, even for
$\rm Pm>100$, we did not find smaller values than $\rm Ha\simeq50$.  This finding is important for objects with very high magnetic Prandtl number like PNS and protogalaxies. They can possess stable toroidal magnetic fields of finite value, up to $\rm Ha\simeq50$.

Obviously, for very small  magnetic Prandtl number the AMRI becomes less
and less important. Magnetic fields with $\rm Ha <Ha_{\rm Tayler}$ are stable
for almost all Re.  For $\rm Ha >Ha_{\rm Tayler}$ this zone of stability
separating the AMRI and TI branches plays an important role though. For a broad range of Reynolds numbers even strong magnetic fields are stabilized against the Tayler instability.

\section{Growth rates}
As we have demonstrated  with Fig. \ref{fig11} (top)  the current-free AMRI shown in Fig. \ref{fig1} has growth rates $\gamma=-\Im({\omega})$ of around
$0.05\Omega_{\rm in}$, only very weakly dependent on Pm.  To characterize the
instabilities with weak currents we compute the complex frequency and the 
wave number in the instability domain shown in Fig. \ref{fig2} for two values of Pm and  for two particular, sufficiently large Reynolds numbers. On 
both the limiting magnetic fields, of course, the growth rate  vanishes. 
Its maximal value between both the limits is (here) 0.004 (see Fig. \ref{fig3}).  That there is a proportionality $\gamma \propto \rm Ha^2$ of the
growth rate    as suggested by Spruit (1999, his Eq. 37) cannot be confirmed. 

\begin{figure}
\hbox{
\includegraphics[width=4cm,height=5cm]{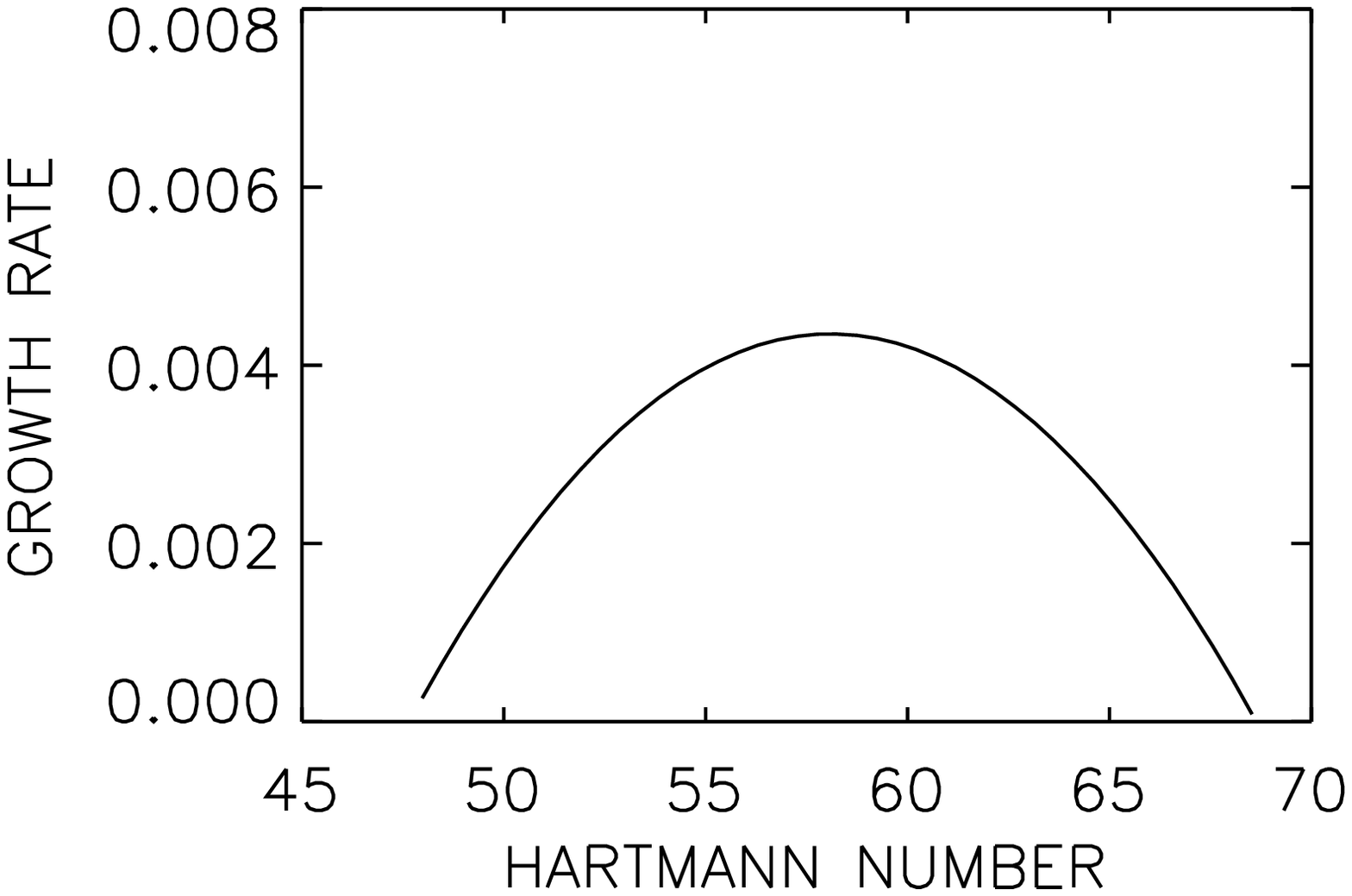}
\includegraphics[width=4cm,height=5cm]{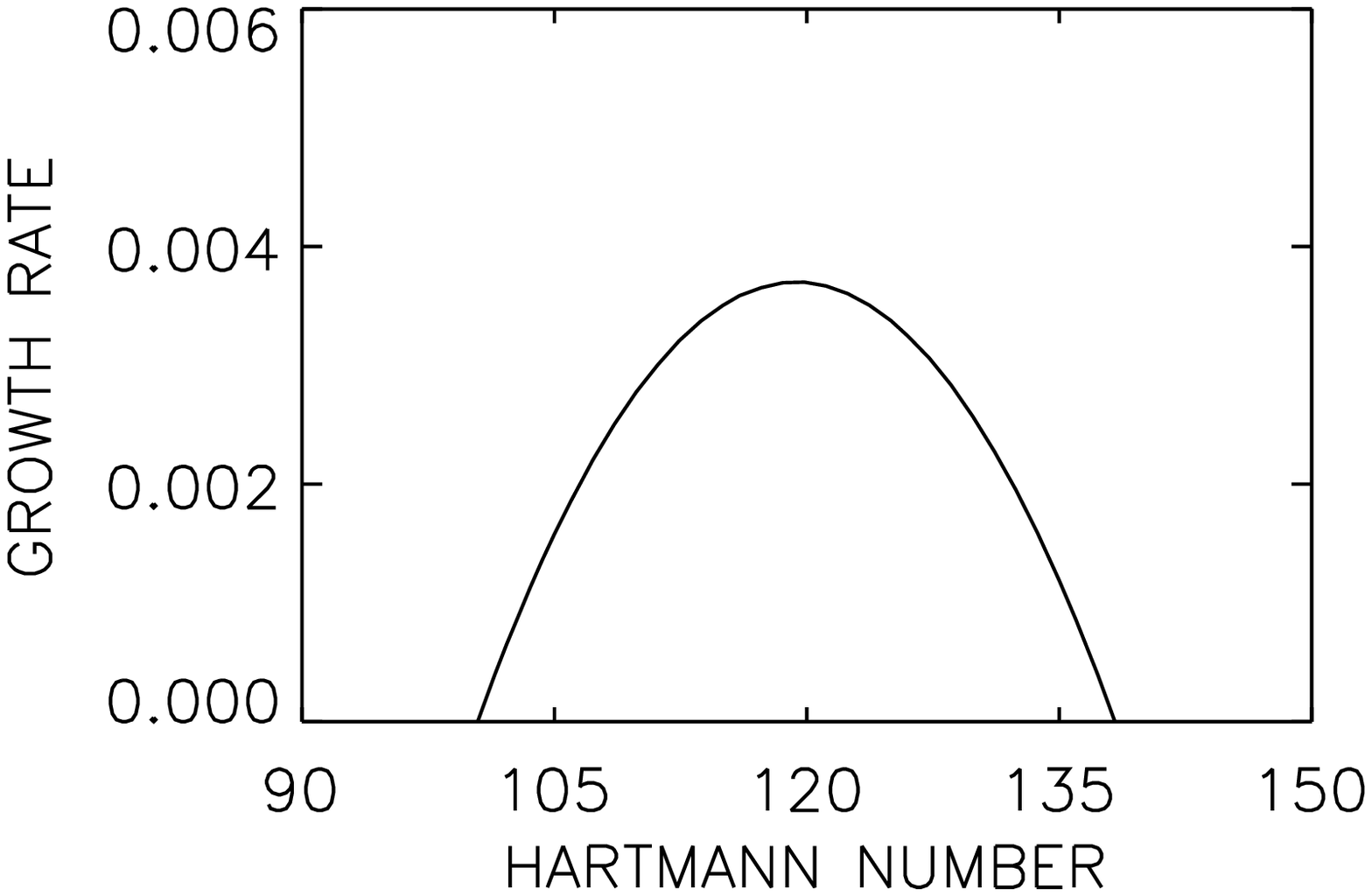}}
\caption{\label{fig3} The growth rate  for the azimuthal magnetorotational instability with weak currents  (see  Fig. \ref{fig2})  for $\rm Pm=1$ (left, here $\rm Rm=125$) and for $\rm Pm=0.1$ (right, here $\rm Rm=82$).}
\end{figure}

The resulting  growth rate of only 0.004 means that the instability grows
with a characteristic time of about 40 rotation periods of the inner cylinder,
i.e. for 20 rotation periods of the outer cylinder. It is a rather slow
instability. However, the growth rates also depend  on the Reynolds number.
They are very small close to the marginal stability limit and they grow with
growing Reynolds number. For slightly supercritical Reynolds numbers one finds 
for ${\rm Pm}=1$ the growth rate expression $\gamma
\simeq(\rm Re-Re_{\rm crit})/1333$ with $\rm Re_{crit}=119$ (Fig. \ref{fig4},
left). The faster the rotation the faster the growth rate, but there is a
saturation of the growth rates, which does not depend strongly on Pm
(see Fig. \ref{fig4}, right). The saturation value  for $\rm Pm=1$ is about
0.12 and for (the more realistic) $\rm Pm=0.1$ it is about 0.10. This maximum
growth rate translates into about 1.5 rotation times of the inner cylinder.
Note therefore that the AMRI is a fast instability for turbulent media with
their typical value of $\rm Pm \leq 1$, but it may be slower for the laminar
gas of radiative stellar cores ($\rm Pm \leq 10^{-{(2\dots3)}}$).
\begin{figure}
\hbox{
\includegraphics[width=4cm,height=5cm]{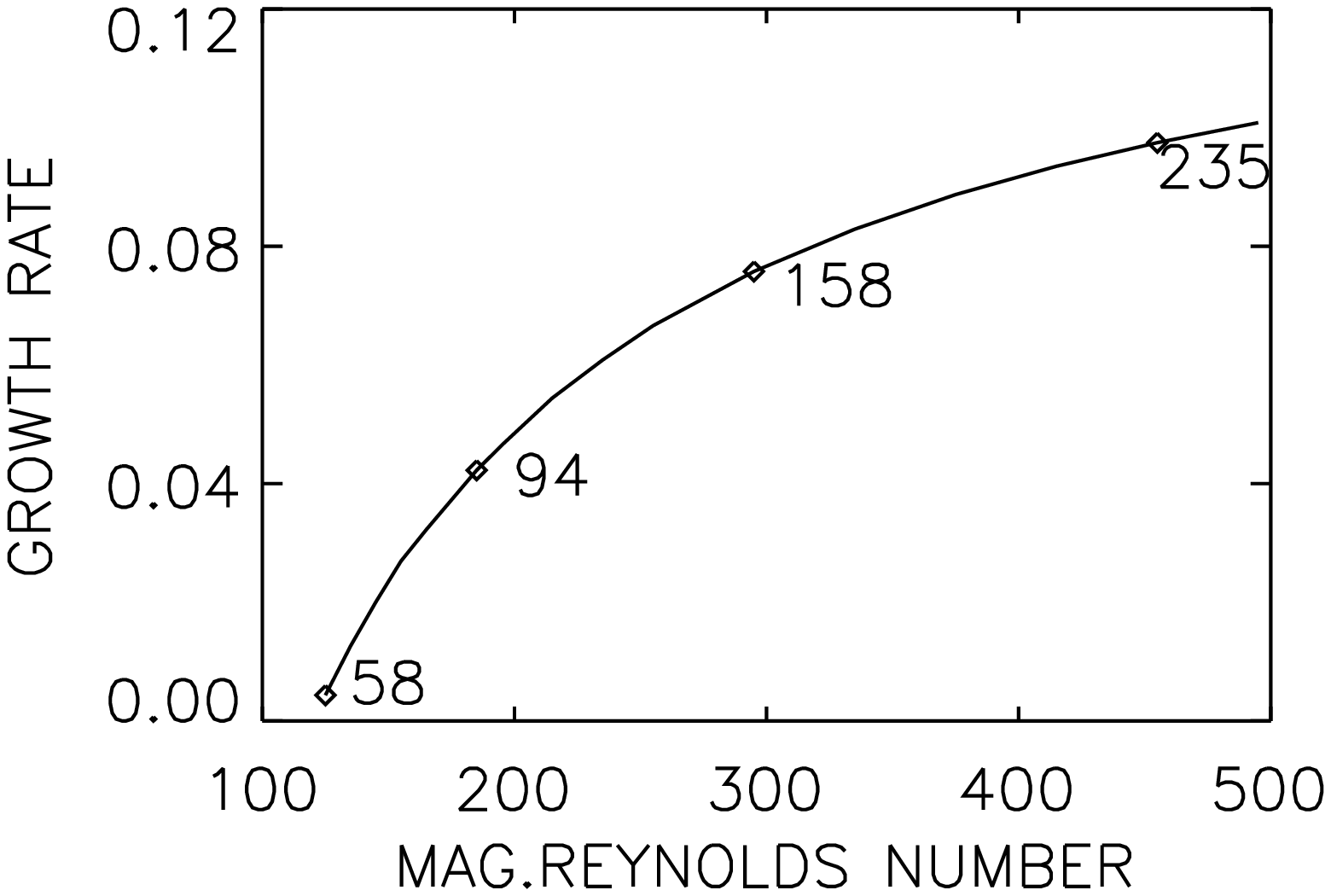}
\includegraphics[width=4cm,height=5cm]{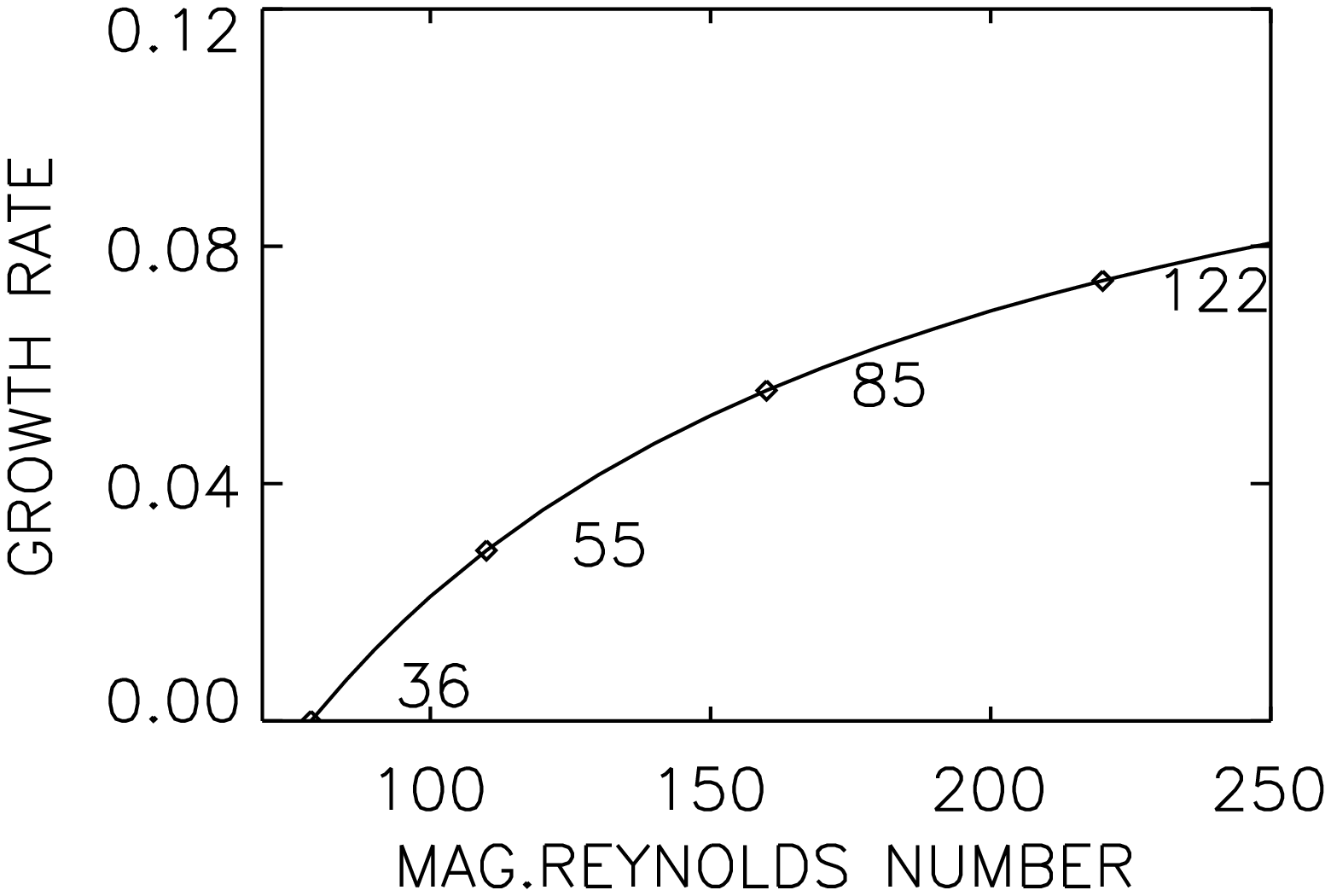}}
\caption{\label{fig4} The growth rate shown in Fig. \ref{fig3} for growing magnetic Reynolds numbers  Rm for $\rm Pm=1$ (left) and $\rm Pm=0.1$ (right).
The curves are marked with the Lundquist  numbers  S for which the growth
rates have a  maximum.}
\end{figure}  
  
The main sequence stars of spectral class A  as a group are fast rotators
with  high Reynolds numbers. The magnetic Ap stars are the slower rotators
in the group so that they should  have  smaller 
Reynolds numbers than the nonmagnetic A stars.  Inspecting Fig. \ref{fig2}
(for small magnetic Prandtl 
number, bottom) one may speculate that the  A stars as a group are located at the limit between AMRI and the stability branch. Then the Ap stars  would lie in the slow-rotation stable regime  while the nonmagnetic A stars would  lie 
in the unstable AMRI regime. Toroidal fields induced by differential rotation of order $10^7$ Gauss would then 
be stable for the slow rotators (Ap) and unstable for the fast rotators (A). It is necessary, 
however, for this kind of theory of the Ap stars that in Fig. \ref{fig2} (bottom) the differential-rotation-induced stable domain 
also exists for sufficiently high Reynolds numbers and Hartmann numbers,
which is still unknown (cf.  Braithwaite \& Spruit 2004).
   
The TI seems to be faster than the AMRI. The growth rates  show that only (say) one rotation time  is enough to develop the instability. 
\begin{figure}
\includegraphics[width=7cm,height=5.0cm]{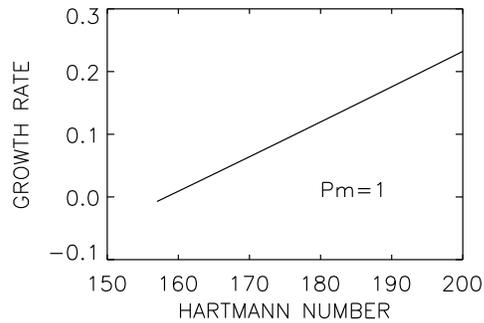}
\caption{\label{fig8} The growth rates  for $\rm Re= 40$ in the TI corner of Fig. \ref{fig2} (middle, $\rm Pm=1$). Note that the growth rates for TI are  higher than for AMRI and that they vary linearly with the magnetic
field.}
\end{figure}
The growth rate presented in Fig. \ref{fig8} scales linearly with the magnetic field. This result clearly confirms the early finding of Goossens, Biront \& Tayler (1981) who found for nonrotating stars with toroidal fields the relation
\beg
\gamma  \gsim\frac{B_\phi }{\sqrt{\mu_0\rho}R}
\label{goos1}
\ende
for the growth rate. Translated to our notation this might mean
\beg
\frac{\gamma}{\Omega} \gsim \frac{\rm Ha}{\rm Re \sqrt{Pm}},
\label{goos2}
\ende
leading to values exceeding unity in accordance to their  numerical results  (their Table 1). For a typical A star Goossens \& Veugelen (1978) find $50\dots 500$ days.
The values in Fig. \ref{fig8} which we obtained under the presence of rotation are of the same order, leading to growth times of one rotation period.   
\section{Azimuthal drift}
For almost all nonaxisymmetric phenomena there is an azimuthal drift of the
pattern $\dot \phi=-\Re(\omega)/m$. Negative (positive)  real parts of the 
frequency $\omega$, therefore, means  eastward (westward) migration.  If the 
pattern is considered in the system corotating with the outer 
cylinder the value 0.5 must be added, which has already been done in Figs.
\ref{fig7} and \ref{fig9}. Then the resulting normalized drift is $-0.05$ in units of the outer rotation, or $-0.025$ in units of the inner rotation. The negative sign 
means that the nonaxisymmetric pattern rotates {\em faster} than the outer cylinder. For different Reynolds numbers  we found very similar drift rates. For increasing Hartmann numbers, however, the effective drift rate is reduced and eventually changes its sign, but the maximal amplitude remains of the order of 0.05. 

The characteristic time for one complete revolution of the pattern is then
about 20 rotation times of the outer cylinder, or 1.35 yrs if we apply these
ideas to the Sun.  This time is well-known from various sorts 
of solar observations (Howe, Komm \& Hill 2002; Krivova \& Solanki 2002; Ternullo 2006). It is also true that sunspots rotate faster than the solar plasma by about 5\%, in agreement with the characteristic timescale of 20 rotation periods. The  timescale of 20 rotation periods seems to form the long-searched timescale governing the cycle time of the activity, intermediate
between the basic timescale of one rotation period and the diffusion timescale.

\begin{figure}
\vbox{
\includegraphics[width=7cm,height=5.0cm]{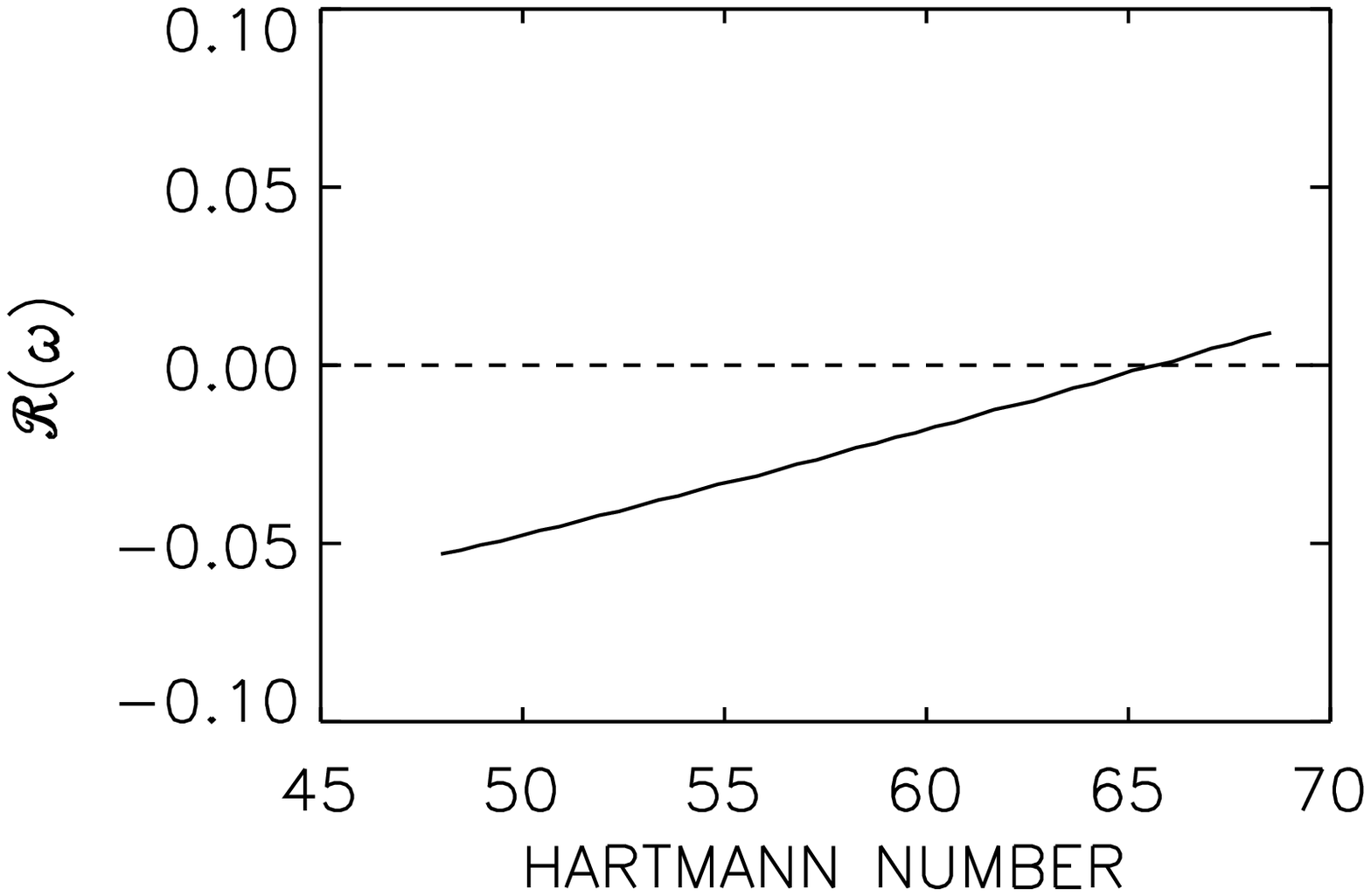}
\includegraphics[width=7cm,height=5.0cm]{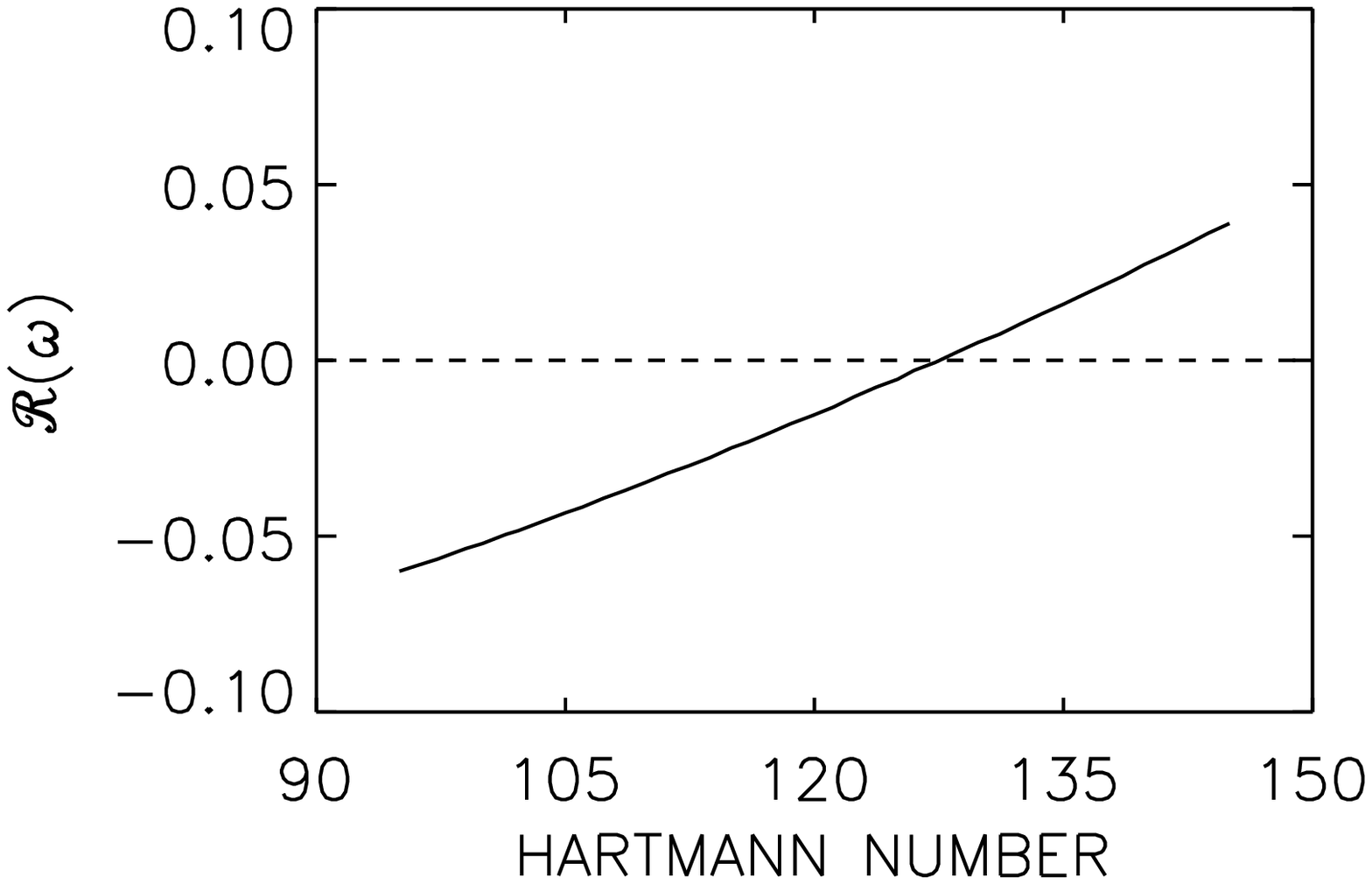}
}
\caption{\label{fig7} The azimuthal drift for $\rm Pm=1$ (top, $\rm Rm=125$) and  $\rm Pm=0.1$ (bottom, $\rm Rm=82$)  hardly depends on the magnetic Prandtl number.  $\hat \mu_\Omega=0.5$.}
\end{figure}

The azimuthal drift velocity of the $m=1$ pattern of the Tayler instability appears to be much faster. The value 0.3 given in Fig. \ref{fig9} does not depend on the field amplitude. It means 4.0 rotation times of the inner cylinder or 2 rotation times of the outer cylinder. This is a strong effect which cannot be missed by the astrophysical observations. If the flip-flop phenomenon of the FK Coma stars is a result of the Tayler instability then the rotation rate of the magnetic pattern should strongly differ from that of  the nonmagnetic stellar plasma. 
\begin{figure}
\includegraphics[width=7cm,height=5.0cm]{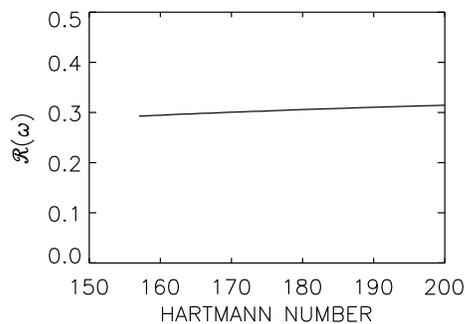}
\caption{\label{fig9} The same as in Fig. \ref{fig8} but for the  drift rate. }
\end{figure}

\section{Discussion}
With a simple global model we have demonstrated how the interaction of
differential rotation and a toroidal magnetic field works. Our main concern is
the presentation of the large influence of the magnetic Prandtl number. The
rotation law considered is stable in the hydrodynamic regime. We 
generally found that a large magnetic Prandtl number is destabilizing, while a
small magnetic Prandtl number has a stabilizing influence, and even
leads to a broad stable domain between the azimuthal magnetorotational
instability (AMRI) and the Tayler instability (TI).

Indeed, for large Pm the instabilities dominate and stable fields can only
exist for  very low Reynolds numbers (Fig. \ref{fig2}, top). Almost always
toroidal fields with $\rm Ha\simeq 50$ are unstable.  For nonturbulent
protoneutron stars (PNS)  one finds critical magnetic fields of only 1 Gauss
and for nonturbulent protogalaxies one finds critical values of only
$10^{-18}$ Gauss. There is no possibility of the existence of stronger
toroidal magnetic fields in nonturbulent PNS and nonturbulent protogalaxies. 

\begin{table*}
\tabcolsep5pt
\caption{\label{tab1} Characteristic values for the solar convection zone  (SCZ) and for a turbulent galaxy.
}
\begin{tabular}{l|cccccc|cccc}
\hline
 & $\rho$ [g/cm$^3$]  & $\nu$ [cm$^2$/s]  & $\eta$ [cm$^2$/s] & $B$ [Gauss] & $\Omega$ [1/s] &$R_0$ [cm]&Re &Rm  &Ha& S\\[0.5ex]
\hline\\[-8pt]
SCZ, top & 10$^{-3}$ &  $10^{13}$ &  $10^{12}$ &$10^{3}$ & $2\cdot 10^{-6}$& $5\cdot 10^{9}$& 5&50 &20& 50\\[0.5ex]
SCZ, bottom & 10$^{-1}$ &  $10^{13}$ &  $10^{13}$ &$10^{4}$ & $2\cdot 10^{-6}$& $10^{10}$& 20& 20& 10& 10\\[0.5ex]
 galaxy & 10$^{-24}$ & $10^{26}$ & $10^{26}$ &$5\cdot  10^{-6}$ &10$^{-15}$&$10^{22}$&1000&1000 &200&200 \\[0.5ex]
    \hline
 \end{tabular}
 \end{table*} 

The opposite is true if turbulent values of the diffusivities are taken into account.
Then  the effective magnetic Prandtl number in cosmical turbulent fields should be 
between 0.1 and 1 (Yousef, Brandenburg \& R\"udiger 2003). In Table \ref{tab1} the 
numerical values are given for the solar convection zone  and for a typical 
galaxy. Note that the  Reynolds number for the galaxy basically  exceeds the solar value. 
For galaxies, therefore, the AMRI should occur for Hartmann numbers of order 50 (see Fig. \ref{fig2}). 
A  value of about 200 is derived from the galactic parameters used  in Table \ref{tab1} for a 
magnetic field of (say) 5 $\mu$Gauss, which represents the observed magnetic fields. Galaxies are 
thus suspected to exist nearby  or within the azimuthal magnetorotational instability  for nonaxisymmetric 
disturbances. The distinct nonaxisymmetric ($m=1$) magnetic geometry of M81 might easily be the result 
of the existence of the presented nonaxisymmetric 
instabilities of azimuthal fields.

For the turbulent solar convection zone  the Reynolds numbers are so small  that 
for $\rm Pm\leq 1$ only the Tayler instability limits the magnetic fields. The related 
Hartmann numbers are about 150. This limit is reached for $B_\phi$ of order 1 kGauss if the dynamo 
operates  in the supergranulation layer (Brandenburg 2005) or for 100 kGauss if it is located in the bulk of the solar convection zone. 

This number is even reduced to only 3 kGauss if -- as it is necessary for the operation of  advection-dominated solar dynamos -- the magnetic diffusivity at the bottom of the convection zone was only $10^{10}$ cm$^2$/s. In this case all azimuthal fields stronger than 3 kGauss become unstable against the nonaxisymmetric Tayler instability
and cannot be amplified any more. Generally,  for too small  magnetic diffusivities (Dikpati \& Gilman, 2006, are using values between $3\cdot 10^{10}$ cm$^2$/s and $2 \cdot 10^{11}$ cm$^2$/s) the Hartman numbers easily grow beyond the critical value and the nonaxisymmetric instabilities -- which do not appear in axisymmetric codes -- basically   limit the magnetic field amplitudes.



\end{document}